# Prediction of topological insulating behavior in $Hg_2CuTi$-type Heusler compounds from first principles

X. M. Zhang,[1] W. H. Wang,[1, a)] E. K. Liu,[1] Z. Y. Liu,[2] G. D. Liu,[3] and G. H. Wu[1]

[1]*Beijing National Laboratory for Condensed Matter Physics, Institute of Physics, Chinese Academy of Sciences, Beijing 100080, P. R. China*

[2] *State Key Laboratory of Metastable Material Sciences and Technology, Yanshan University Technology, Qinhuangdao 066004, P. R. China*

[3] *School of Material Sciences and Engineering, Hebei University Technology, Tianjin 300130, P. R. China*

Abstract:

The topological band structures of the $X_2YZ$ Heusler compounds with the $Hg_2CuTi$ structure are investigated by using first-principles calculations within density functional theory. Our results clearly show that a large number of the $Hg_2CuTi$ type Heusler compounds naturally exhibit distinct band-inversion feature, which is mainly controlled by the Y-Z zinc blende sublattice. Similar to the half-Heusler family, the topological band order in $Hg_2CuTi$ type Heusler compounds is sensitive to the variation of lattice constant, and most of them possess a negative formation energy, which makes them more suitable in material growth and could easily achieve the topological insulating behavior by alloying or proper strain.

---

a) Electronic mail: wenhong.wang@iphy.ac.cn

# I. INTRODUCTION

Topological insulator (TI) with a full insulating gap in the bulk generated by strong spin-orbit coupling (SOC), but contains conducting states on edges or surfaces, has been a hot topic within the condensed matter physics community during its discovery in recent years. [1-4] The topologically protected gapless surface states are chiral and inherently robust to external perturbations, so that the TI is meaningful for future technological applications in spintronics and quantum computing as well. [5,6]

Since the first discovery of a two-dimensional TI in an HgTe based quantum well,[6,7] several other families of materials for three-dimensional (3D) TI have been proposed theoretically and recently studied experimentally.[8-16] For example, tetradymite semiconcuctors, such as $Bi_2Te_3$, $Bi_2Se_3$, and $Sb_2Te_3$ are confirmed to be 3D TI with a single Dirac-cone on the surface, where the bulk band gap is as large as 0.3 eV, making the room temperature application possible.[8-12] However, a clear shortcoming of $Bi_2Te_3$ family is that these materials cannot be made with coexisting magnetism, a much desired property for spintronic applications. Although doping can be used to achieve the magnetically ordered behavior,[17] this creates extra complexity in material growth and could introduce detrimental effects upon doping.

Assisted with topological band theory, first-principles calculations have played an important role in uncovering new families of TIs. [7-13] Since the first two-dimensional (2D) TI with quantum spin-Hall effect was predicted in HgTe, [14-16] many families of materials have been proposed to be three-dimensional (3D) TIs from binary compounds such as $Bi_{1-x}Sb_x$, $Bi_2Te_3$, $Bi_2Se_3$ and $Sb_2Te_3$, [17-19] then a number of the rare earth containing half-Heusler compounds have been predicted to be topologically nontrivial. [12,20] Very recently, the investigations on TIs have also gained great success experimentally: not only in binary compounds, but also the gapless half-Heusler compounds has been confirmed to possess metallic surface electronic state factually. [21,22] Most importantly, it was proposed that in the half-Heusler family the topological insulator allows the incorporation of superconductivity and/or magnetism. [23]

In particular, the earlier computations (Refs.11 and 12) viewed the structure of the XYZ Heusler compounds as $X^{n+}$ "stuffed" $YZ^{n-}$ zinc blende. Moreover, the details of the band structure near Fermi level in Half-Heusler compounds exhibit striking similarity with those of CdTe and HgTe, one reason of which is that Half-Heusler and zinc blende lattice possess the same crystal symmetry with space group $F\bar{4}3m$ (No 216). Cannot be ignored, in Full-Heusler compounds, Whose stoichiometric composition is $X_2YZ$, there exists a family of materials sharing the $F\bar{4}3m$ crystal symmetry as well, called $Hg_2CuTi$-type Heusler compounds. In particular, we have made a comparison of $Hg_2CuTi$ and zinc-blende structures in Fig.1. The former can be looked upon as four interpenetrating face-centered-cubic (fcc) lattices, in which the X atoms occupy A (0,0,0) and B(1/4, 1/4, 1/4) sites, the Y atom at C (1/2,1/2,1/2) and Z at D (3/4,3/4,3/4) in Wyckoff coordinates. In a similar way, the zinc-blende structure consists of two interpenetrating fcc lattices sitting the nearest site C and D. Just make an analogous consideration with the previous studies, [11,12] the $Hg_2CuTi$ structure thus can be considered as two hybridized interpenetrating zinc blendes, which are X-X and Y-Z. Owning to many compounds in zinc blende and Half-Heusler structure with 18 valence electrons have been expected to exhibit a gap at the Fermi energy and some are proved topologically nontrivial as well. [14-16, 20] It is therefore very important to study whether the topologically nontrivial phase is developed in $Hg_2CuTi$-type Heusler compounds with 18 valence electrons.

In this letter, by performing a systematic investigation of the band topology of some given $Hg_2CuTi$ type Heusler compounds based on first-principles calculations, we show that a large number of potential TIs are waiting for exploit in this rich $Hg_2CuTi$ type Heusler family. Our deduction and theoretical basis are mostly judged by making a comparison with the topologically nontrivial binary compound HgTe, [14-16] whose band structures has been thoroughly studied and can be looked upon as a benchmark. Similar to the half-Heusler family, we also found that the topological insulating behavior in these $Hg_2CuTi$ Heusler compounds is sensitive to the variation of lattice constant and tetragonal uniaxial strain as well.

## II. COMPUTATIONAL DETAILS

To investigate the band topology, we employ the full-potential linearized augmented plane-wave method, [24] implemented in the package WIEN2K.[25] Experimental lattice constants are used when available, and others are obtained by minimizing the total energy using generalized gradient approximation (GGA) of Perdew-Burke-Ernzerhof 96 including SOC. [24, 26] A converged ground state was obtained using 5 000 k points in the first Brillouin zone and $K_{max}* R_{MT} = 8.0$, where $R_{MT}$ represents the muffin-tin radius and $K_{max}$ the maximum size of the reciprocal-lattice vectors. The muffin-tin radius of the atoms used in caculations are generated by the system automatically. Moreover, wave functions and potentials inside the atomic sphere are expanded in spherical harmonics up to l=10 and 4, respectively.

### III. RESULTS AND DISCUSSIONS

As we have mentioned before, the $Hg_2CuTi$ type Heusler compounds with 18 valence electrons are supposed to exhibit analogous band structure with CdTe/HgTe, since the $Hg_2CuTi$ structure can be considered as two interpenetrating zinc blende sublattices and share the same crystal symmetry with CdTe and HgTe. We thus provide a typical example which compares the calculated band structures of CdTe and HgTe with those of $Y_2RuPb$ and $Sc_2OsPb$. The results are shown in Fig.2. For clarity, we have marked the relevant bands with different colors in our work. Just divide into two groups for comparison: CdTe [Fig.2 (a)] and $Y_2RuPb$ [Fig.2(b)], HgTe [Fig.2(c)] and $Sc_2OsPb$ [Fig.2(d)]. Obviously, the band structures of the two groups seem extremely alike respectively. Firstly, the band structure detail of $Y_2RuPb$ is very similar to that of CdTe with natural band ordering [*s*-like $\Gamma_6$ states (red lines) lie above the *p*-like $\Gamma_8$ states (blue lines)] and open a direct gap at the $\Gamma$ point, which indicate them just be trivial semiconductors. However, the compounds in the other group namely HgTe and $Sc_2OsPb$ possess the same inverted band order, in which the $\Gamma_6$ state is occupied and sits below the $\Gamma_8$ state. At the same time, the valence and conduction bands away from the $\Gamma$ point are well separated without crossing each other. That is to say the band inversion only occurs once throughout the Brillouin zone and therefore, HgTe and $Sc_2OsPb$ are both topologically nontrivial phases in their ground states.

Armed with the above analysis, next we perform a systematic investigation of the band topology of the $X_2YZ$ (X = Sc, Y, La; Y=Ru, Re, Os; Z= Sb, Pb, Bi) compounds, which are all proposed to crystallize in $Hg_2CuTi$ structure. Considering the band structures of $Hg_2CuTi$ type Heusler compounds are extremely similar to CdTe/HgTe and the low-energy electron dynamics is dominated by energy bands at the $\Gamma$ point, the topology of $Hg_2CuTi$ type Heusler compounds can also be characterized by a similar band inversion of $\Gamma_6$ and $\Gamma_8$ levels. Here we define $\Delta E = E_{\Gamma_6} - E_{\Gamma_8}$ as topological band inversion strength (TBIS), which would be positive for topologically trivial cases and negative for topologically nontrivial phases. Fig.3 (a) shows $\Delta E$ as a function of the lattice constant for the calculated $Hg_2CuTi$ type Heusler compounds, including the cases of CdTe and HgTe for comparison. It can be seen that most of the materials exhibit a negative $\Delta E$, which indicate them TI candidates, such as $Sc_2OsPb$ and $Y_2OsPb$. While there are also some cases like $Sc_2ReSb$ and $La_2OsPb$ that the Fermi level visibly cut the conduction or the valence bands, though have a negative $\Delta E$, we can only call them topological metals. Regarding $Y_2RuPb$, $Sc_2RuPb$ and $La_2RuPb$, they are trivial semiconductors with a positive $\Delta E$ like CdTe. While the $Hg_2CuTi$ structure can be considered as two hybrided interpenetrating zinc blende sublattices: X-X and Y-Z. It is clear enough that, when the Y-Z zinc blende is Ru-Pb, the $Hg_2CuTi$ Heusler type compounds will have a positive $\Delta E$, no matter varying composition of X-X zinc blende. Inversely, if the Y-Z zinc blendes are Os-Pb, Re-Bi or Re-Sb, the compounds are prone to be topologically nontrivial. That is to say, the sign of ΔE is dominated by the Y-Z zinc blende and X-X just contributes as a fine tuning of the value only. The reasons can be easily illustrated as follows: the atoms in Y-Z zinc blende are much heavier than those in X-X and introduce a dominant drastic interplay of the SOC which plays a vital role in the band inversion mechanism. This can be fully supported by the previous studies. [11,12,20]

In order to investigate whether the $Hg_2CuTi$ type Heusler compounds can be naturally synthesized in experiment, we have calculated the formation energy using the formula: $E_b=(E_{tol} - m^* E_X - n^* E_Y - p^* E_z)$, in which the $E_{tol}$ is the energy of the $Hg_2CuTi$ Heusler alloys under equilibrium lattice constants, and $E_X$, $E_Y$ and $E_Z$ are the energy when X/Y/Z crystallized in pure metals, respectively. The coefficients m, n, p show proportions of X, Y, Z in the $Hg_2CuTi$ formula, which are 2, 1, 1 respectively. In Fig.3 (b), we show the formation energy of the $Hg_2CuTi$ type Heusler compounds as a function of the lattice constant. The result shows that all of the calculated Heusler compounds have a negative formation energy, and some values are even lower than that of HgTe and CdTe, which have already been successfully synthesized and well-studied experimentally. [14-16] From the view of the calculated formation energy, we can conclude that these $Hg_2CuTi$ type Heusler compounds studied in our work are promising to be synthesized for experimental characterization.

In fact that the topological insulating behavior of half-Heusler compounds not only influenced by interplay of the SOC (mainly produced by heavy atoms), but the degree of hybridization (controlled by the lattice constant). [11,12] Thus the trivial insulators can be converted to topological nontrivial phases by applying proper hydrostatic expansion and vice versa. In our Hg$_2$CuTi type Heusler compounds, for example, a 1.8% change in the lattice constant converts the trivial insulator Y$_2$RuPb [Fig.4 (a)] into a nontrivial topological insulator [Fig.4 (b)]. Conversely, the Sc$_2$OsPb [Fig.4(c)] with inverted band order in its native state will transform into a trivial topological phase [Fig.4 (d)] under a 2.1% hydrostatic compression. These behaviors are quite similar to those in Half-Heusler and chalcopyrite family. [11-13, 20] However, the topological band order in Hg$_2$CuTi type Heusler compounds appears more sensitive to the variation of lattice constant according our calculations: nearly all the Hg$_2$CuTi type Heusler compounds studied in our work can make a band order turn under less than 3% lattice distortion, which may be meaningful in practice TIs design and application.

Similar to zinc blende and half-Heusler families, these topologically nontrivial Hg$_2$CuTi type Heusler compounds are not naturally insulating because the two couples of bands with $\Gamma_8$ symmetry degenerating together at the $\Gamma$ point owning to the protection of cubic symmetry. However, the degeneracy between the subbands of $\Gamma_8$ symmetry can be broken and opened a direct gap by proper strain or doping engineer, [9-11, 20] which drives the zero-gap semiconductor into a real TI phase. In order to demonstrate this mechanism in our Hg$_2$CuTi type Heusler compounds, we apply a ± 5% uniaxial strain along the [001] direction with constant volume to Sc$_2$OsPb system. The results are shown in Fig.5. It can be seen that, as a result of reducing the cubic crystal symmetry into tetragonal, the fourfold degeneracy of the $\Gamma_8$ states is broken and formed a gable band gap indeed. Interestingly, the system exhibits different responses upon the uniaxial compression and expansion: when introduced a compression [Fig.5 (a)], the conduction and valence bands are no longer overlapped and the system becomes an insulator with inverted band order retained, while upon expansion [Fig.5 (b)] it remains a semimetal as before. The result is well consistent with our previous calculations for half-Heusler LaPtBi.[28] Indeed, a recent search model for possible TIs has suggested that the variational 'descriptor', namely 'strain' can be associated with the robustness or the feasibility of the TI state.[29]

## IV. CONCLUSION

In summary we have shown by *ab-initio* calculations that a large number of $Hg_2CuTi$ Heuslers of composition $X_2YZ$ (X = Sc, Y, La; Y=Ru, Re, Os; Z= Sb, Pb, Bi) exhibits an inverted band order naturally and is promising to realize the topological insulating order. Their band structures are determined by X-X and Y-Z zinc blende sublattices jointly, in which the Y-Z dominates the sign of topological band inversion strength $\Delta E$ and X-X just contributes as a fine tuning. We also found that the band topology is sensitive to the variation of lattice constant and uniaxial strain, which is consistent with those half-Heusler families. Importantly, most of $Hg_2CuTi$ type Heusler compounds possess a negative formation energy making them more suitable in material growth and could easily achieve the topological insulating behavior by proper strain. With the example of $Sc_2OsPb$, we show how the gapless system can be driven by uniaxial strain along the [001] direction into a topological insulating state.

**Acknowledgements**

This work was supported by National Natural Science Foundation of China (Grant Nos. 51171207, 51021061 and 51025103) and National Basic Research Program of China (973 Programs: 2012CB619405).

**References:**

[1] H. Zhang, C.-X. Liu, X.-L. Qi, X. Dai, Z. Fang, and S.-C. Zhang, Nat. Phys. 5, 438 (2009).

[2] X.-L. Qi and S.-C. Zhang, Phys. Today 63, No. 1, 33 (2010).

[3] J. E. Moore, Nature 464, 194 (2010).

[4] M. Z. Hasan, C. L. Kane, Rev. Mod. Phys. 82, 3045 (2010).

[5] J. E. Moore, Nat. Phys. 5, 378 (2009).

[6] L. Fu, C. L. Kane, and E. J. Mele, Phys. Rev. Lett. 98, 106803 (2007).

[7] J. E. Moore and L. Balents, Phys. Rev. B 75, 121306 (2007).

[8] R. Roy, Phys. Rev. B 79, 195322 (2009).

[9] W. Al-Sawai, H. Lin, R. S. Markiewicz, L. A. Wray, Y. Xia, S. Y. Xu, M. Z. Hasan, and A. Bansil, Phys. Rev. B 82, 125208 (2010).

[10] W. X. Feng, D. Xiao, Y. Zhang, and Y. G. Yao, Phys. Rev. B 82, 235121 (2010).

[11] S. Chadov, X.-L. Qi, J. Kübler, G. H. Fecher, C. Felser, and S.-C. Zhang, Nature Materials 9, 541 (2010).

[12] H.-J. Zhang, S. Chadov, L. Muchler, B. Yan, X.-L. Qi, J. Kübler, S.-C. Zhang, and C. Felser, Phys. Rev. Lett. 106, 156402 (2011).

[13] W. Feng, D. Xiao, J. Ding, and Y. Yao, Phys. Rev. Lett. 106, 016402 (2011).

[14] B. A. Bernevig, T. L.Tughes and S.-C. Zhang, Science, 314, 1757 (2006).

[15] M. Konig et al., Science 318, 766 (2007).

[16] M. Konig, S. Wiedmann, C. Brüne, A. Roth, H. Buhmann, L. Molenkamp, X. -L. Qi, and S. -C. Zhang, Science, 318, 766 (2007).

[17] D. Hsieh, D. Qian, L. Wray, Y. Xia, Y. S. Hor, R. J. Cava, and M. Z. Hasan, Nature 452, 970 (2008).

[18] Y. Xia, D. Qian, D. Hsieh, L. Wray, A. Pal, H. Lin, A. Bansil, D. Grauer, Y. S. Hor, R. J. Cava, and M. Z. Hasan, Nat. Phys. 5, 398 (2009).

[19] Y. L. Chen, J. G. Analytis, J.-H. Chu, Z. K. Liu, S.-K.Mo, X. L. Qi, H. J. Zhang, D. H. Lu, X. Dai, Z. Fang, S. C. Zhang, I. R. Fisher, Z. Hussain, and Z.-X. Shen, Science 325, 178 (2009).

[20] H. Lin, L. A. Wray, Y. Xia, S. Xu, S. Jia, R. J. Cava, A. Bansil, and M. Z. Hasan, Nature Materials 9, 546 (2010).

[21] D. Xiao et al., Phys. Rev. Lett. 105, 096404 (2010).

[22] D. Hsieh, J.W. McIver, D. H. Torchinsky, D. R. Gardner, Y. S. Lee, and N. Gedik, Phys. Rev. Lett, 106, 057401 (2011).

[23] C. Liu, Y.B. Lee, T. Kondo, E. D. Mun, M. Caudle, B. N. Harmon, Sergey L. Bud'ko, P. C. Canfield, A. Kaminski, Phys. Rev. B 83, 205133 (2011).

[24] N. P. Butch, P. Syers, K. Kirshenbaum, J. Paglione, Phys. Rev. B 84, 220504(R) (2011).

[25] D. J. Singh, Plane waves, Pseudopotentials and the LAPW Method (Kluwer Academic, Boston, 1994).

[26] P. Blaha, K. Schwarz, G. Madsen, D. Kvaniscka, and J. Luitz, Wien2k, An Augmented Plane Wave Plus Local Orbitals Program for Calculating Crystal Properties (Vienna University of Technology, Vienna, Austria, 2001).

[27] J. P. Perdew, K. Burke, and M. Ernzerhof, Phys. Rev. Lett. 77, 3865 (1996).

[28] X. M. Zhang, W. H. Wang, E. K. Liu, G. D. Liu, Z.Y. Liu Y, G. H. Wu, Appl. Phys. Lett. 99 071901 (2011)

**Figure captions:**

FIG. 1. (Color online) Comparison of the $Hg_2CuTi$ and zinc-blende crystal structures. The $Hg_2CuTi$ ($X_2YZ$) and the zinc-blende (YZ) structures are shown in (a) and (b), respectively.

FIG. 2. (Color online) Band structures of CdTe and HgTe compared with $Hg_2CuTi$-type $Y_2RuPb$ and $Sc_2OsPb$ Heusler compounds. The $\Gamma_6$ and $\Gamma_8$ states are denoted by red and blue lines, respectively. This comparison reveals obvious similarity between the two systems: both CdTe and $Y_2RuPb$ are trivial semiconductors with $\Gamma_6$ situated above $\Gamma_8$. However, both HgTe and $Sc_2OsPb$ are topological nontrivial with inverted band order only.

FIG. 3. (Color online) (a) Energy difference between $\Gamma_6$ and $\Gamma_8$ bands ($\Delta E = E_{\Gamma 6} - E_{\Gamma 8}$) and (b) formation energy of the calculated $Hg_2CuTi$ type Heusler compounds as a function of the lattice constant. Here HgTe and CdTe binaries are shown for comparison.

FIG. 4. （Color online） Band structures of $Y_2RuPb$ and $Sc_2OsPb$. $Y_2RuPb$ (a) without and (b) with a 1.8% hydrostatic expansion, $Sc_2OsPb$ (c) without and (d) with a 2.0% hydrostatic compression. Here, the red and blue lines represent the subbands with $\Gamma_6$ and $\Gamma_8$ symmetry, respectively. The application of a hydrostatic expansion in $Y_2RuPb$ causes the $\Gamma_6$ states to jump below the $\Gamma_8$ states, and leads to a nontrivial topological phase. In contrast, the application of a hydrostatic compression in $Sc_2OsPb$ causes the $\Gamma_6$ states to jump above the $\Gamma_8$ states, and leads to a trivial topological phase.

FIG. 5. （Color online） Band structure of $Sc_2OsPb$ under uniaxial strain with constant volume along [001] direction, with a reduction the c/a ratio by 5% in (a), and an increase in the c/a ratio by 5% in (b).

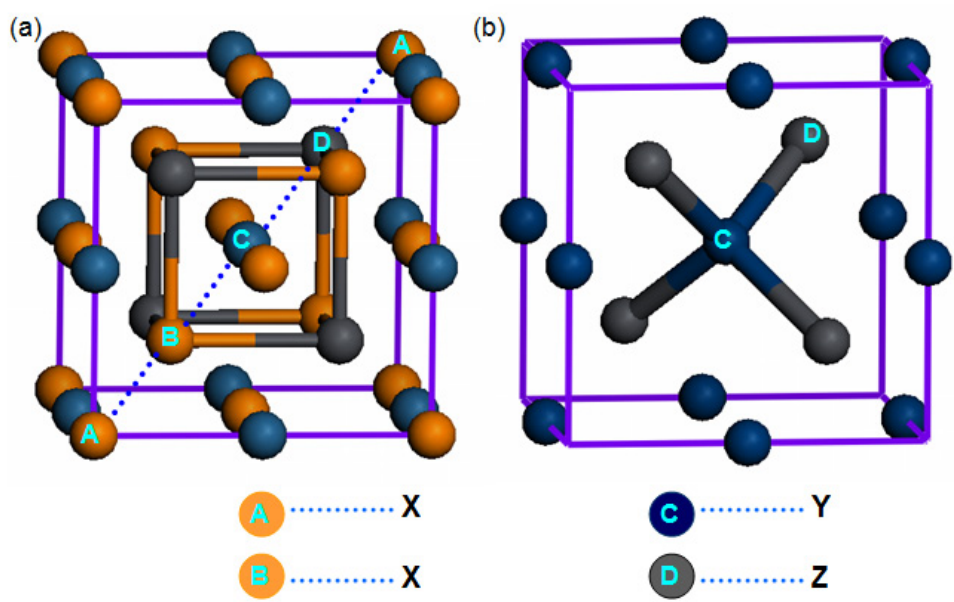

FIG. 1. Zhang et al.,

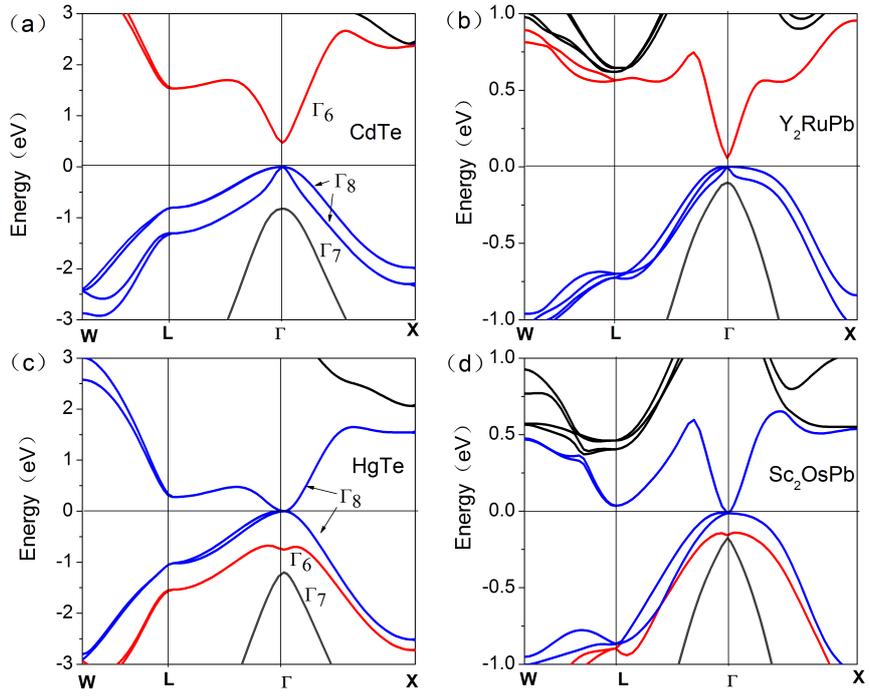

FIG. 2. Zhang et al.,

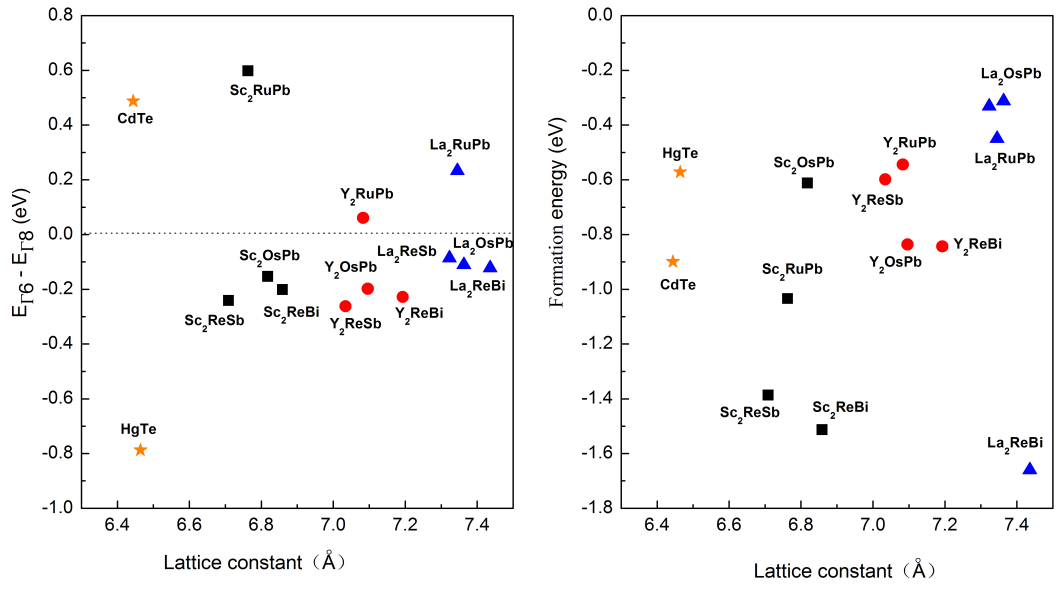

FIG. 3. Zhang et al.,

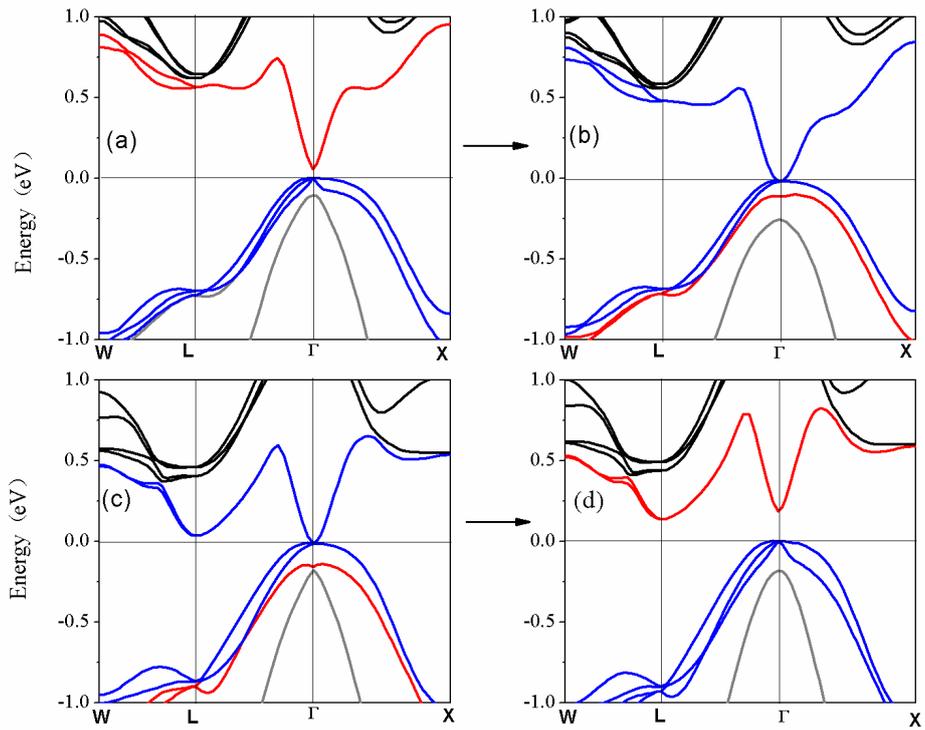

FIG. 4. Zhang et al.,

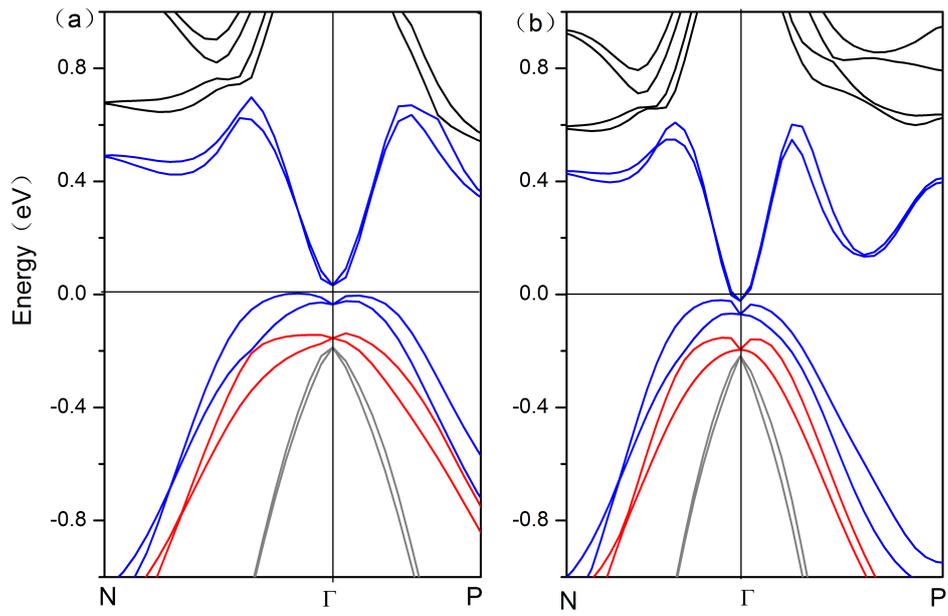

FIG. 5. Zhang et al.,